\documentclass[aps,english,showpacs,twocolumn,pra]{revtex4-1}
\usepackage{amsfonts}
\usepackage{amssymb}
\usepackage{amsmath}
\usepackage{graphicx}
\usepackage{epsfig}

\begin{document}

\title{Semi-localization transition driven by a single asymmetrical tunneling%
}
\author{P. Wang}
\author{K. L. Zhang}
\author{Z. Song}
\email{songtc@nankai.edu.cn}
\affiliation{School of Physics, Nankai University, Tianjin 300071, China}
\begin{abstract}
A local impurity usually only strongly affects few single-particle energy
levels, thus cannot induce a quantum phase transition (QPT), or any
macroscopic quantum phenomena in a many-body system within the Hermitian
regime. However, it may happen for a non-Hermitian impurity. We investigate
the many-body ground state property of a one-dimensional tight-binding ring
with an embedded single asymmetrical dimer based on exact solutions. We
introduce the concept of semi-localization state to describe a new quantum
phase, which is a crossover from extended to localized state. The peculiar
feature is that the decay length is of the order of the system size, rather
than fixed as a usual localized state. In addition, the spectral
statistics is non-analytic as asymmetrical hopping strengths vary, resulting
a sudden charge of the ground state. The distinguishing feature of such a
QPT is that the density of ground state energy varies smoothly due to
unbroken symmetry. However, there are other observables, such as the
groundstate center of mass and average current, exhibit the behavior of
second-order QPT. This behavior stems from time-reversal symmetry
breaking of macroscopic number of single-particle eigen states.
\end{abstract}

\maketitle

\section{Introduction}

Understanding the quantum phase transitions (QPTs) is of central
significance to both condensed matter physics and quantum information
science. QPTs occur only at zero temperature due to the competition between
different parameters describing the interactions of the system. A
quantitative characterization of a QPT is that certain quantity, such as
order parameter and Chern number undergoes qualitative changes when some
parameters pass through quantum critical points. So far almost all the
investigations about QPT focus on systems with translational symmetry, in
aid of which the local order parameter and topological invariant can be well
defined. In both cases, the groundstate property is encoded in complete set
of single-particle eigenstates, forming Bogoliubov quasiparticle band or
Bloch band. A conventional symmetry-breaking QPT concerns all the
single-particle eigenstates independently, regardless of the connection
between them, while a topological QPT captures global features of the
symmetry-respecting single-particle eigenstate sets. On the other hand, the
translational symmetry indicates that the QPT is driven by a global
parameter, such as external field or uniform coupling constant. There are
two prototypical exactly solvable models, transverse-field Ising model \cite%
{SachdevBook} and QWZ model \cite{QWZ}, based on which the concept and
characteristic of\ conventional and topological QPTs can be well
demonstrated.

Intuitively, the translational symmetry is not necessary for the onset of a
QPT, a material in practice usually has an open boundary condition. A
fundamental question is whether QPTs can be driven by a local parameter.
However,\ a local parameter usually only strongly affects few
single-particle energy levels, thus cannot induce a QPT, or any macroscopic
quantum phenomena in a many-body system within the Hermitian regime. It is
well known that a non-Hermitian system may make many things possible,
including quantum phase transition that induces in a finite system \cite%
{Znojil1,Znojil2,Bendix,LonghiPRL,LonghiPRB1,Jin1,Znojil3,LonghiPRB2,LonghiPRB3,Jin2,Joglekar1,Znojil4,Znojil5,Zhong,Drissi,Joglekar2,Scott1,Joglekar3,Scott2,Tony}%
, unidirectional propagation and anomalous transport \cite%
{LonghiPRL,Kulishov,LonghiOL,Lin,Regensburger,Eichelkraut,Feng,Peng,Chang},
invisible defects \cite{LonghiPRA2010,Della,ZXZ}, coherent absorption \cite%
{Sun} and self sustained emission \cite%
{Mostafazadeh,LonghiSUS,ZXZSUS,Longhi2015,LXQ}, loss-induced revival of
lasing \cite{PengScience}, as well as laser-mode selection \cite%
{FengScience,Hodaei,JLPRL}. Such kinds of novel phenomena can be traced to
the existence of exceptional point, which is a transition point of symmetry
breaking for a pair of energy levels. Exploring novel quantum phase or QPT
\cite{LCPRA1,LCPRA2,LCPRA3,LCPRB,WRPRA,LSPRB,JLPRB,ZXZPRA,ZKLPRB} in
non-Hermitian systems becomes an attractive topic. Motivated by the recent
development of non-Hermitian quantum mechanics \cite{CMBender}, both in
theoretical and experimental aspects \cite%
{PRL08a,PRL08b,Klaiman,CERuter,YDChong,Regensburger,LFeng,Fleury,BenderRPP,NM,FL,Ganainy18,YFChen,Christodoulides}%
, in this paper we investigate the QPTs in non-Hermitian regime. The purpose
of the present work is to present a simple non-Hermitian model to
demonstrate alternative type of QPT, which driven by a local parameter. We
study a phenomenon that we dub semi-localization, which is induced by a
single asymmetrical tunneling embedded in a uniform tight-binding ring. A
semi-localization state is a crossover from extended to localized states,
possessing a truncated exponentially decay probability distribution. The
peculiar feature is that the decay length is of the order of the size of the
system, rather than fixed as usual localized state. The single-particle
solution of the model shows that the spectral statistics, such as the number
and distribution of the complex energy levels, is controlled by the
asymmetrical hopping strength. The eigenstate is a semi-localized state for
a complex level, while an extended state for a real level. Particularly, a
real (complex)-level wave function possesses symmetry (asymmetric)
probability distributions and steady (non-steady) with zero (nonzero)
current due to the unbroken (broken) time-reversal symmetry. Although the
system is non-Hermitian with complex single-particle spectrum, the many-body
groundstate energy is always real due to the protection of time reversal
symmetry. It exhibits a unconventional QPT arising from the sudden change of
the single-particle spectral statistics: The density of many-body
groundstate energy is analytic, while the center of mass and average
staggered current of the ground state, as macroscopic quantities, are
non-analytic functions of the asymmetric hopping strength. Accordingly, the
transition from fully real to complex spectrum is associated with the
transition from extension to semi-localization.

This paper is organized as follows. In Sec. \ref{Model and solution}, we
present a non-Hermitian time-reversal symmetric model with asymmetric dimer
and the Bethe Ansatz solution. In Sec. \ref{Phase diagram}, we provide the
phase diagram by analysing the properties of eigenstates with real and
complex energy levels, such as the proportion of complex level, the
groundstate center of mass, and the groundstate average staggered current.
In Sec. \ref{Phase transition}, we demonstrate the characteristics of
second-order QPT. Finally, we give a summary in Section \ref{Summary}.

\section{Model and solution}

\label{Model and solution}

\begin{figure}[tbp]
\centering
\includegraphics[width=0.48\textwidth]{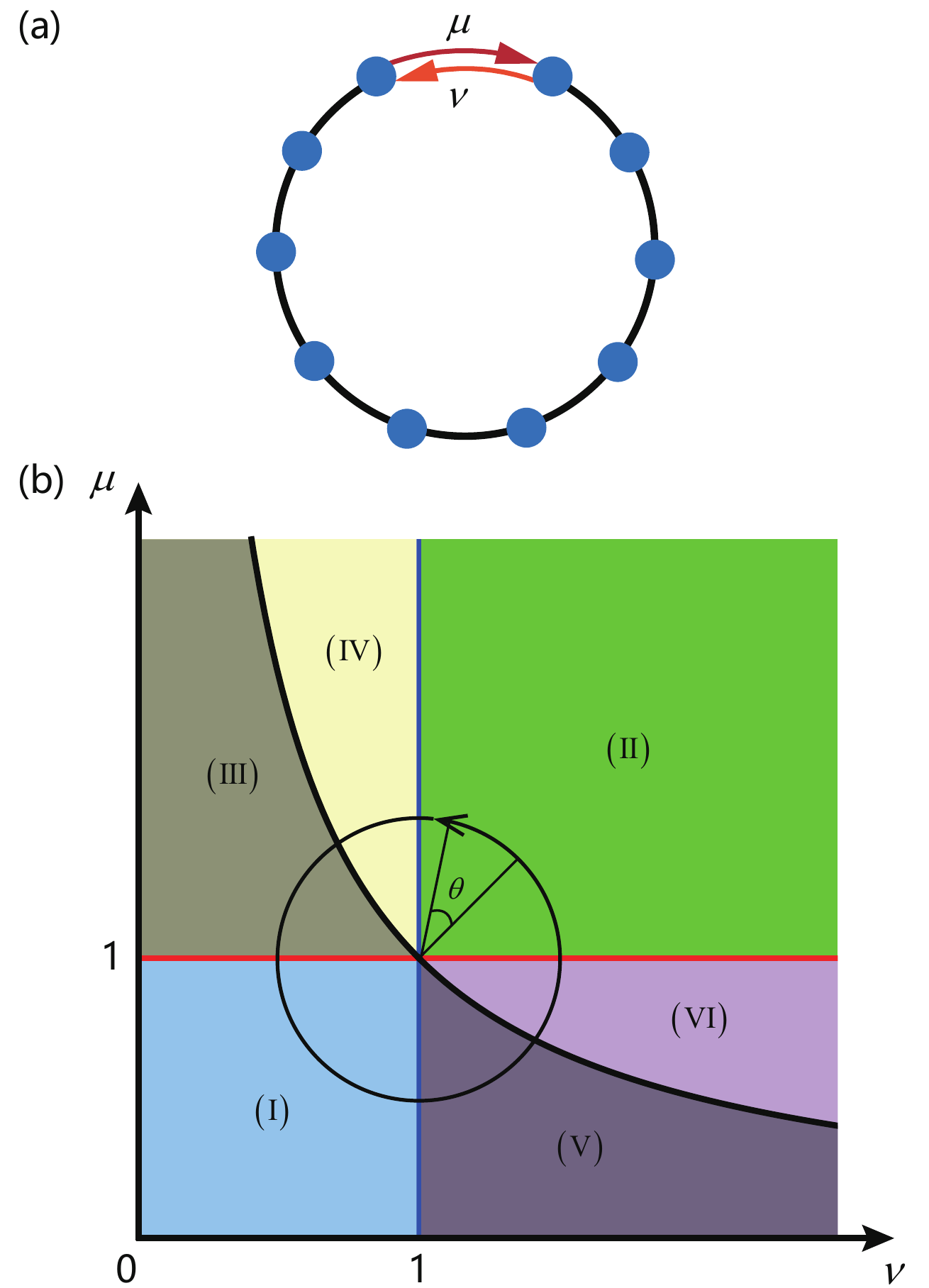}
\caption{(a) Schematic of the model, which depicts a uniform tight-binding
ring with unitary hopping strength, embedded by a single non-Hermitian
dimer. It is an asymmetric tunneling with hopping strength $\protect\mu $\
and $\protect\nu $. (b) Phase diagram of the system, which is consisted of
six regions. In regions (I) and (II), system has full real spectrum, while
in (III), (IV), (V) and (VI) the spectra are mixed with real and complex
energy levels. At curve $\protect\mu \protect\nu =1$, all the energy levels
are complex. The inset circle described by parameter equations $\protect\nu %
=1+r\cos (\protect\theta )$; $\protect\mu =1+r\sin (\protect\theta )$, goes
through all six regions and will be used in the following figures.}
\label{fig1}
\end{figure}

\begin{figure*}[tbh]
\centering
\includegraphics[width=1\textwidth]{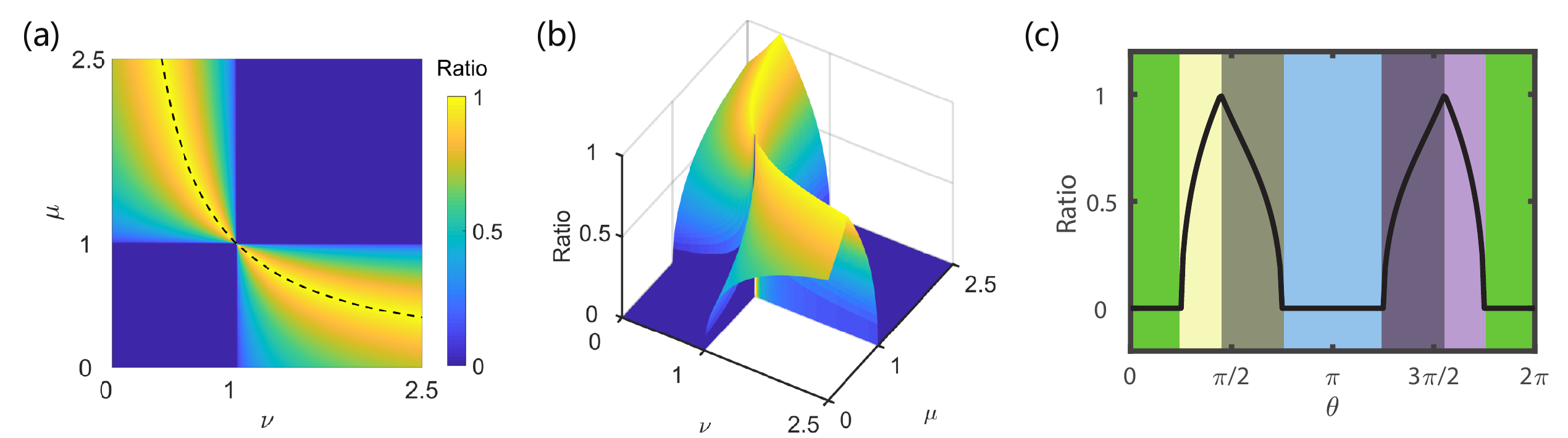}
\caption{Plots of the ratio$\ g(\protect\mu ,\protect\nu )$ in $\protect\nu
\protect\mu $-plane from Eq. (\protect\ref{g}), as one of perspectives of
the spectral statistics. (a) Color contour plot. The dished lines indicates
the curve $\protect\nu \protect\mu =1$. The-nonanalytic behavior of $g(%
\protect\mu ,\protect\nu )$ at lines $\protect\mu \protect\nu =1$\ are
obvious.\ (b) 3D plot. The nonanalytic behavior of $g(\protect\mu ,\protect%
\nu )$ at curve $\protect\mu \protect\nu =1$\ is highlighted. (c) Plots of
the ratio at the loop $\protect\nu =1+0.5\cos (\protect\theta )$; $\protect%
\mu =1+0.5\sin (\protect\theta )$. The nonanalytic behaviors at the phase
boundaries are clearly indicated by the sharp peaks and right angle turns.}
\label{fig2}
\end{figure*}

Considering a simple uniform tight-binding ring, it is well known that the
spectrum is cosine type and cannot be changed largely by a local impurity in
general. An additional Hermitian hopping term or even non-Hermitian local
on-site complex potential can only alter several energy levels, introducing
localized states. However, we will see that another type of non-Hermitian
impurity may have an affect on macroscopic energy levels, which plays the
key role in the present work.

The Hamiltonian has the form%
\begin{equation}
H=\sum\limits_{j=1}^{N-1}c_{j}^{\dag }c_{j+1}+\mathrm{H.c.}+\mu c_{N}^{\dag
}c_{1}+\nu c_{1}^{\dag }c_{N},
\end{equation}%
with odd $N/2$, where $c_{j}$ is the annihilation operator of fermion at
site $j$.\ It depicts a uniform tight-binding ring with only a non-Hermitian
impurity embedded. A schematic illustration of the model is presented in
Fig. \ref{fig1}(a). The non-Hermiticity arises from an asymmetric tunneling
between sites $1$ and $N$, represented by hopping strength $\mu $\ and $\nu $
(in this paper, we only consider the case with $\mu ,$\ $\nu >0$ for
simplicity). This model is investigated in the previous work \cite{WPPRA} in
the special case with $\mu \nu =1$. It has been shown that an asymmetric
dimer can be realized by the combination of imaginary potential and magnetic
flux \cite{LCPRA}. Experiments on asymmetric dimer has been proposed\textbf{%
\ }\cite{Longhi,LFengNC,ZGong}\textbf{.}

Unlike usual many-body non-Hermitian tight-binding model, Hamiltonian $H$
does not have parity-time symmetry and translational symmetry. Owing to the
reality of the coupling $\mu $\ and $\nu $, it possesses time reversal
symmetry, ie., $\left[ \mathcal{T},H\right] =0$, where $\mathcal{T}$\ is an
anti-unitary operator with action $\mathcal{T}^{-1}i\mathcal{T}=-i$.
Fortunately, the solution of $H$ can be exactly obtained by the Bethe ansatz
technique (see Appendix).

The Hamiltonian can be diagonalized as the form%
\begin{equation}
H=\sum\limits_{n=1}^{N}\varepsilon _{n}\overline{\gamma }_{n}\gamma _{n}
\end{equation}%
where the fermion operators $\overline{\gamma }_{n}$ and $\gamma _{n}$\ have
the form%
\begin{equation}
\overline{\gamma }_{n}=\sum_{l=1}^{N}f_{n}^{l}c_{l}^{\dag },\gamma
_{n}=\sum_{l=1}^{N}\overline{f_{n}^{l}}c_{l},
\end{equation}%
and satisfy the canonical commutative relation%
\begin{equation*}
\left\{ \gamma _{m},\overline{\gamma }_{n}\right\} =\delta _{mn}.
\end{equation*}%
Here the canonical conjugate operator can be constructed by the relation%
\begin{equation}
\overline{f_{n}^{l}}(\mu ,\nu )=\left[ f_{n}^{l}(\nu ,\mu )\right] ^{\ast }
\end{equation}%
and the explicit expression of wave function is%
\begin{equation}
\left\{
\begin{array}{c}
f_{n}^{l}=\frac{1}{\sqrt{\Omega }}\sin \left( k_{n}l+\alpha _{n}\right)
,(2<l<N-1) \\
f_{n}^{1}=\frac{1}{\sqrt{\Omega }},f_{n}^{N}=\frac{e^{-ik_{n}}-\mu
e^{ik_{n}\left( N-1\right) }}{\nu -e^{ik_{n}N}}f_{1}%
\end{array}%
\right.
\end{equation}%
where $\alpha _{n}$ is obtained by $\tan \alpha _{n}=c_{n}/s_{n}$, and%
\begin{eqnarray}
s_{n} &=&\nu ^{2}+1-\nu \cos \left( k_{n}N\right) +\nu \cos \left[
k_{n}\left( N+2\right) \right]  \notag \\
&&-p_{n}\cos k_{n}, \\
c_{n} &=&\nu \sin \left( k_{n}N\right) -\nu \sin \left[ k_{n}\left(
N+2\right) \right] +p_{n}\sin k_{n}, \\
p_{n} &=&\cos \left[ k_{n}\left( N+1\right) \right] -\mu \nu \cos \left[
k_{n}\left( N-1\right) \right]  \notag \\
&&+\left( \mu +\nu \right) \cos k_{n}.
\end{eqnarray}%
The coefficient $\Omega _{n}$\ is determined by the biorthonormal inner
product. In the rest of paper, we focus on the Dirac probability, since it
can be measured directly in experiment. Then we take the Dirac normalization
factor which is obtained from $\sum_{l=1}^{N}|f_{n}^{l}|^{2}=1$. The
single-particle spectrum has the form%
\begin{equation}
\varepsilon _{n}=e^{ik_{n}}+e^{-ik_{n}},
\end{equation}%
where $k_{n}$\ can be real and complex. The quasi-wave vector $k_{n}$\ for $%
1\leqslant n\leqslant N$ has the form%
\begin{equation}
k_{n}=\frac{2n\pi }{N}+\theta _{n},
\end{equation}%
where $\theta _{n}$\ is determined by the transcendental equations (see
Appendix). The transcendental equation is reduced to%
\begin{equation}
\tan \theta _{n}=\frac{\left( 1-\mu \nu \right) \sin \left( \theta
_{n}N\right) }{\mu +\nu -\left( 1+\mu \nu \right) \cos \left( \theta
_{n}N\right) },
\end{equation}%
for $n=N$ or $N/2$; and%
\begin{equation}
\left\{
\begin{array}{c}
\sin (\phi _{n}+\theta _{n}N)=\frac{(\mu +\nu )\sin \frac{n\pi }{N}}{\sqrt{%
1+\mu ^{2}\nu ^{2}-2\mu \nu \cos \frac{2n\pi }{N}}} \\
\tan \phi _{n}=\frac{1+\mu \nu }{1-\mu \nu }\tan \frac{n\pi }{N}%
\end{array}%
\right.
\end{equation}%
otherwise. Obviously, the reality of $k_{n}$\ ($1\leqslant n\leqslant N$)\
depends on the values of $\mu $\ and $\nu $, which will be discussed in
detail in the next section.

\section{Phase diagram}

\label{Phase diagram} In this section, we analyze the property of the
solution and the corresponding implications. At first, we determine the
phase diagram from the perspective of spectral statistics, which is
characterized by the proportion of the complex levels. Secondly, we
introduce a concept, semi-localized state, to describe the feature of the
eigenstates of complex energy levels. Furthermore, we reveal another
exclusive property of the complex-level eigenstates, the non-steady, which
only can be seen from an evolved (non-equilibrium) state in a Hermitian
system.

\begin{figure}[tbp]
\centering
\includegraphics[width=0.4\textwidth]{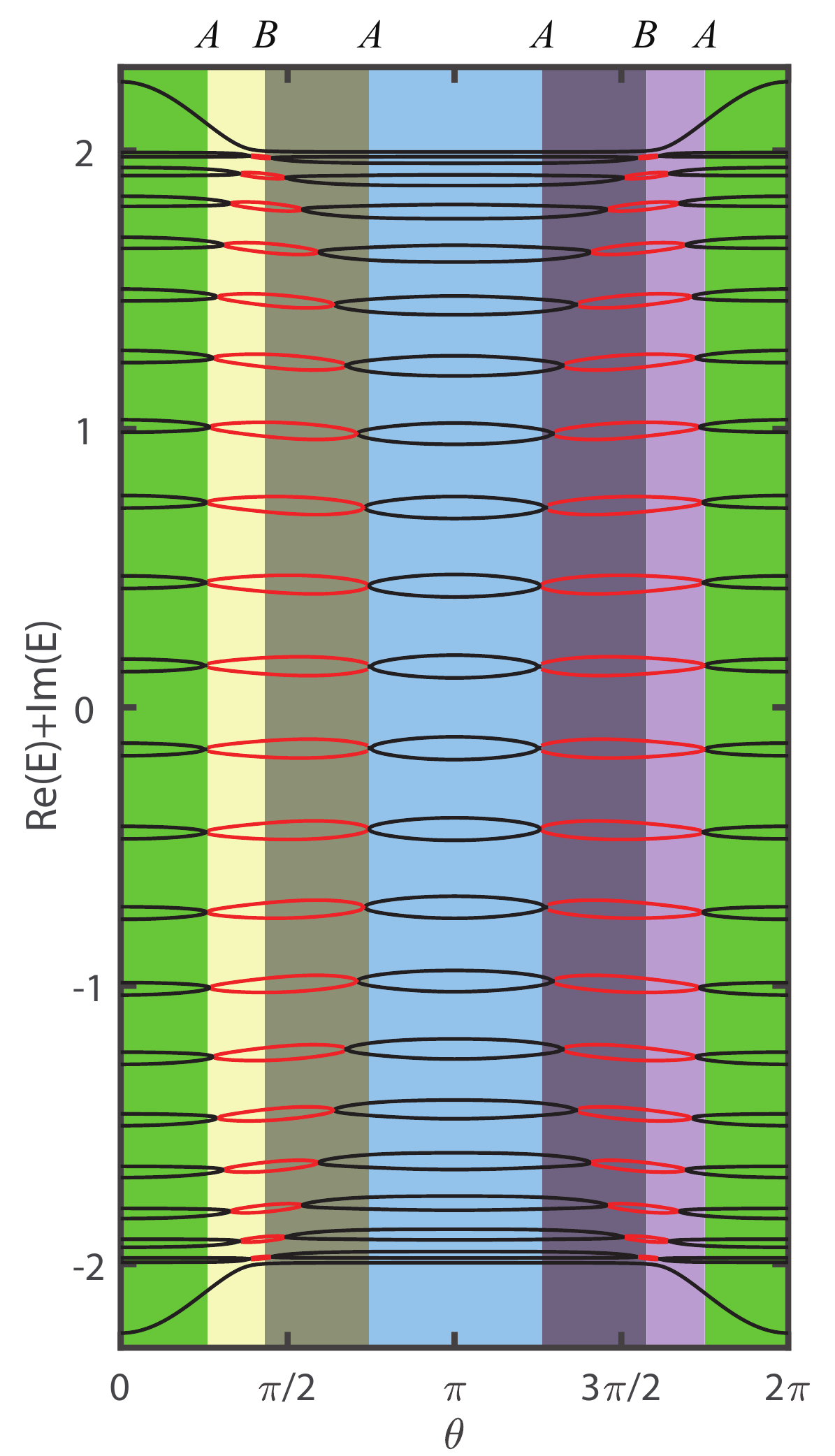}
\caption{Plots of energy level structures for understanding the connection
between the spectral statistics and QPTs. Energy levels for the system on
the loop $\protect\nu =1+0.9\cos (\protect\theta )$; $\protect\mu =1+0.9\sin
(\protect\theta )$ are plotted. Black and red lines represent the real and
complex energy levels, respectively. There are two types of QPTs with the
boundaries indicated by $A$ and $B$, respectively. We see that $A$-type
boundary always corresponds to the appearance of complex levels, while $B$%
-type boundary locates at the maximal number of complex levels. The size of
the system is $N=42$.}
\label{fig3}
\end{figure}

\begin{figure}[tbp]
\centering
\includegraphics[width=0.48\textwidth]{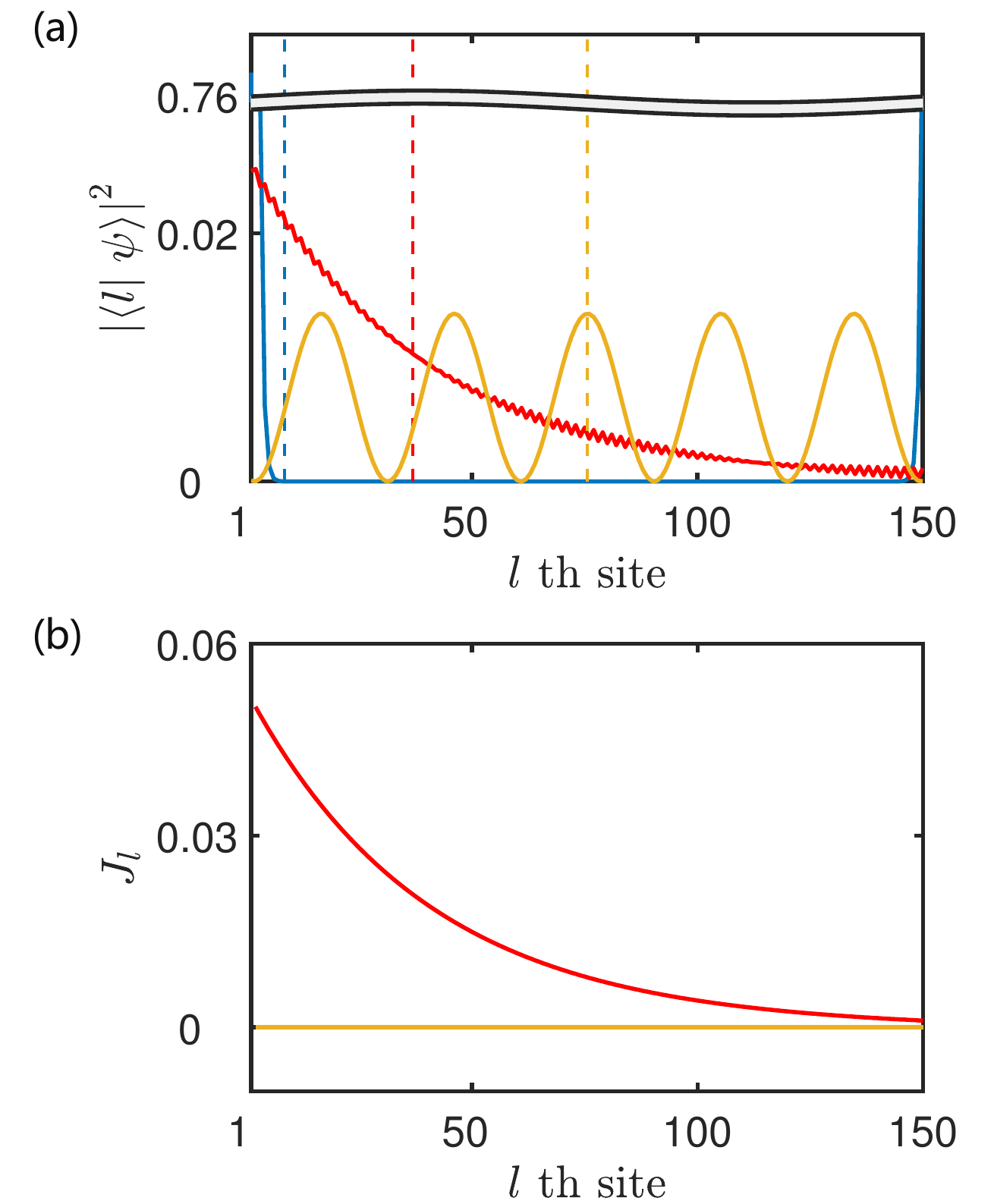}
\caption{Schematic illustrations for the concept of semi-localized state.
(a) Plots of the profiles of three kinds of wave functions: localized state
(blue), extended state (yellow), and semi-localized state (red). The decay
length of the semi-localized state has the same order of the lattice length.
The dash lines represent the value of center of mass for each states. We see
that a semi-localized state is a crossover between localized and extended
states. (b) Plots of the probability current distributions for three kinds
of states. The current for extended state and local state are both zero. The
current distribution of a semi-localized state exhibits a non-steady
behavior, violating the probability conservation, which is exclusive for the
eigenstate of a non-Hermitian system. The size of the system is $N=150$.
Other parameters are $\protect\nu =5$, $\protect\mu =10$ for the blue lines,
$\protect\nu =5$, $\protect\mu =0.1$ for the red lines and $\protect\nu =4$,
$\protect\mu =2$ for the yellow lines.}
\label{fig4}
\end{figure}

\subsection{Spectral statistics}

According to the solutions obtained in the Appendix, the reality of energy
levels obeys the following rules. (i) $\mu ,\nu >1$, or $\mu ,\nu <1$, all
the quasi-wave vectors $k_{n}$ are either real or imaginary, corresponding
to real energy levels. All the eigenstates are non-degeneracy except the
trivial case with $\mu =\nu =1$, which reduces the system to be a uniform
Hermitian ring. We denote the (non-degeneracy) real-energy single-particle
eigenstate as $\left\vert \psi _{\mathrm{R}}^{n}\right\rangle $, barring the
energy levels $n=N$ and $N/2$. (ii) $\mu >1>\nu $, or $\mu <1<\nu $, some
complex quasi-wave vectors $k_{n}$ appear, corresponding to complex energy
levels, which come in pair with conjugate eigen energy. We denote the
complex-energy single-particle eigenstate as $\left\vert \psi _{\mathrm{\pm }%
}^{n}\right\rangle $. (iii) Among them, especially in the case of $\mu \nu =1
$, all $k_{n}$ becomes complex. Obviously, as one of the characteristics of
the spectral statistics, the proportion of the complex level is defined as
function of $\mu $\ and $\nu $%
\begin{equation}
g(\mu ,\nu )=\frac{N_{\mathrm{C}}(\mu ,\nu )}{N},
\end{equation}%
which is the ratio of the number of the complex levels $N_{\mathrm{C}}$ to
the total number of levels. In large $N$ limit, we have%
\begin{equation}
g(\mu ,\nu )=\left\{
\begin{array}{cc}
0, & \mu ,\nu \geqslant 1,\text{or }\mu ,\nu \leqslant 1 \\
(\pi -2k_{c})/\pi , & \mu >1>\nu ,\text{or }\mu <1<\nu  \\
1, & \mu \nu =1%
\end{array}%
\right. ,  \label{g}
\end{equation}%
where\textbf{\ }$k_{c}=|\arcsin [(1-\mu \nu )/(\nu -\mu )]|$ is the critical
wave vector separating the real and complex levels. A schematic of the three
kinds of regions, which will be shown as phase diagram is plotted in Fig. %
\ref{fig1}(b). To demonstrate the properties of the ratio, $3$D profiles of $%
g(\mu ,\nu )$\ and the corresponding energy-level structure are plotted in
Fig. \ref{fig2}(a, b, c) and Fig. \ref{fig3}. We note that $g(\mu ,\nu )$\
is non-analytic at the curves $\mu =1$, $\nu =1$\ and $\mu \nu =1$. We will
show that such curves\ are phase boundaries for many-body ground state due
to the sudden change of the spectral statistics.

\subsection{Semi-localized state}

Unlike a linear operator such as the parity, time reversal operator $%
\mathcal{T}$\ is an anti-linear operator. The $\mathcal{T}$-symmetry
breaking is always associated with the appearance of complex levels. Exact
solution in Appendix shows that the single-particle eigen function can
always be expressed to obey the relations%
\begin{equation}
\mathcal{T}\left\vert \psi _{\mathrm{R}}^{n}\right\rangle =\left\vert \psi _{%
\mathrm{R}}^{n}\right\rangle ,\text{ }\mathcal{T}\left\vert \psi _{\mathrm{%
\pm }}^{n}\right\rangle =\left\vert \psi _{\mathrm{\mp }}^{n}\right\rangle .
\end{equation}%
Owing to the value of $k_{n}$, there are three types of wave functions:
extended, localized and semi-localized states. Here the last one is
exclusive for non-Hermitian system. Unlike the localized one, the imaginary
part of $k_{n}$ of semi-localized states,\ $\theta _{n}$\textbf{\ }is
inversely proportional to\textbf{\ }$N$, leading to an incomplete decay
distribution (with a truncated tail). Nevertheless, it still supports
imbalanced probability distribution, as a crossover from extended to
localized states. We employ the center of mass (CoM), which is the
expectation value of the CoM operator%
\begin{equation}
r_{\mathrm{c}}=\frac{1}{N}\sum\limits_{l=1}^{N}lc_{l}^{\dag }c_{l}.
\end{equation}%
Straightforward derivation shows that (i) $\left\langle r_{\mathrm{c}%
}\right\rangle \approx 1/2$\ for an extended state; (ii) $\left\langle r_{%
\mathrm{c}}\right\rangle \approx 0$\ ($1$) for a localized state at the case
$\nu >\mu $\ ($\mu >\nu $); and (iii) for a semi-localized state $%
\left\langle r_{\mathrm{c}}\right\rangle $\ is a number ranging from $0$\ to
$1$\textbf{.} Here we give an example, when $\mu \nu =1$ and $\nu >1>\mu $,
we have%
\begin{equation}
\left\langle r_{c}\right\rangle \approx \frac{1}{(\sqrt[N]{\nu ^{2}}-1)N}+%
\frac{1}{1-\nu ^{2}},
\end{equation}%
in large $N$ limit. It is easy to check that $\lim_{\nu \rightarrow
1}\left\langle r_{c}\right\rangle =1/2$, which accords with the above
analysis. To demonstrate the above conclusions, profiles of such three types
of states and the corresponding $\left\langle r_{\mathrm{c}}\right\rangle $\
are plotted in Fig. \ref{fig4}a. We can see that the difference among the
three types of eigenstates is obvious.

\subsection{Non-steady eigenstate}

We note that the $\mathcal{T}$-symmetry breaking\ of\ $\left\vert \psi _{%
\mathrm{\pm }}^{n}\right\rangle $ indicates that $\left\vert \psi _{\mathrm{%
\pm }}^{n}\right\rangle $ can have non-zero local current, which is defined
as
\begin{equation}
J_{l}=-i\left\langle \left( c_{l}^{\dag }c_{l+1}-\mathrm{H.c.}\right)
\right\rangle _{n},
\end{equation}%
at the position $l$. Here $\left\langle ...\right\rangle _{n}$\ denotes the
expectation value for an eigenstate of $n$ level. It is usual for a
Hermitian system, for instance, taking $\mu =\nu =1$, each eigenstate with
nonzero momentum has zero local-current. Remarkably, an intriguing feature
is that $\left\vert \psi _{\mathrm{\pm }}^{n}\right\rangle $\ is a
non-steady state, since $J_{l}$\ is position-dependent, violating the
conservation of current. A non-steady state can exist in a Hermitian system,
such as a moving wavepacket in a tight-binding ring. However, it cannot be
an eigenstate of a Hermitian system. If $\mu \leftrightarrows \nu $, the
current changes the sign. In Fig. \ref{fig4}b, profiles of the current
distributions for three types of eigenstates are plotted.

As a temporary summary, we can conclude that a semi-localized eigenstate has
distinguishing feature from an extended one (The localized eigenstate can be
negligible, since there are only two such eigenstates at most). This should
result in macroscopic property for a many-body ground state.

\section{Phase transition}

\label{Phase transition}

\begin{figure*}[tbh]
\centering
\includegraphics[width=1\textwidth]{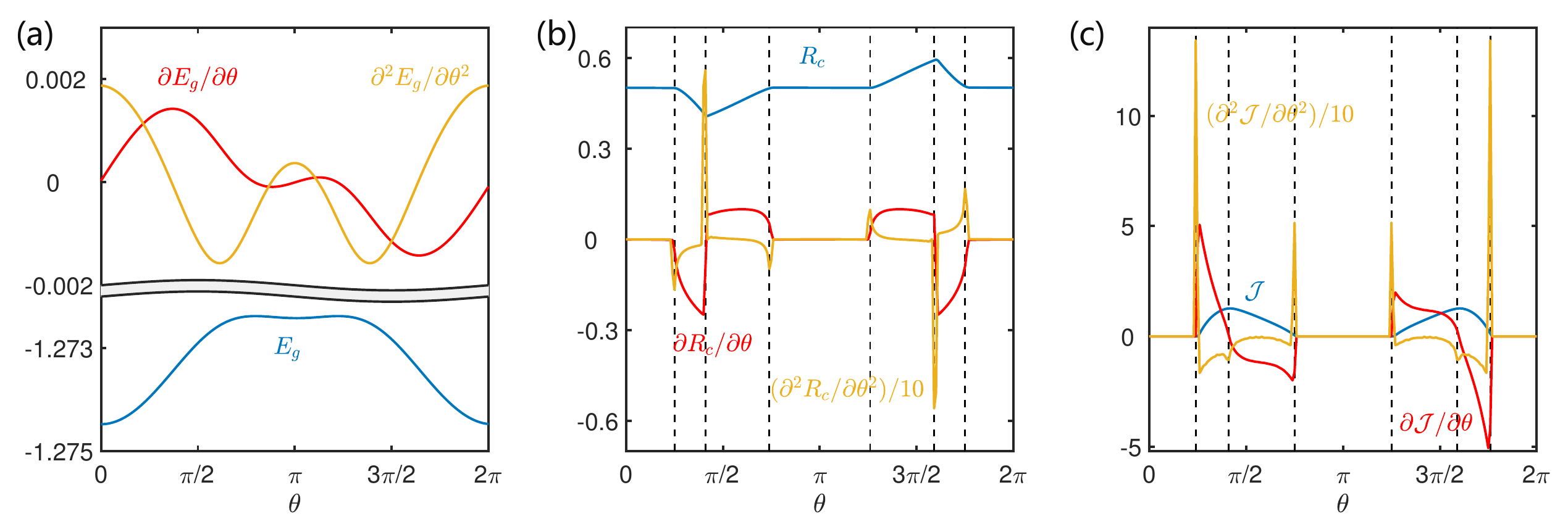}
\caption{Plots of three quantities (a) density of ground state energy, (b)
average CoM. (c) average staggered current, and the corresponding first-,
second-order derivatives as function of $\protect\theta $. It indicates that
the density of groundstate energy does not display any critical behaviors,
while the other two exhibit the characteristics of second-order QPT:
first-order derivatives are non-analytical and second-order derivatives are
divergent. The parameters are $\protect\nu =1+0.9\cos (\protect\theta )$; $%
\protect\mu =1+0.9\sin (\protect\theta )$. The size of the system is $N=750$%
. }
\label{fig5}
\end{figure*}

Now we consider the many-body effect of the single-particle spectral
statistics. We focus on the ground state for half-filled case, where all the
negative real-parts of energy levels are filled by fermions. It is expected
that the non-analyticity of $g(\mu ,\nu )$\ can result in macroscopic
phenomena.

First of all, we consider the density of ground state energy, which is
expressed as%
\begin{equation}
E_{g}=\frac{2}{N}\sum\limits_{n=1}^{N/2}\varepsilon _{n}=\frac{2}{N}%
\sum\limits_{n=1}^{N/2}\text{\textrm{Re}}(\varepsilon _{n}).
\end{equation}%
From the exact result in the Appendix, $E_{g}$\ is always analytical at all
range of $\left\{ \mu ,\nu \right\} $, which seems to indicate that there is
no occurrence of conventional QPT\textbf{.} Secondly, we investigate the
average CoM, which is defined as%
\begin{equation}
R_{\mathrm{c}}=\frac{2}{N}\sum\limits_{n=1}^{N/2}\left\langle r_{\mathrm{c}%
}\right\rangle _{n}.
\end{equation}%
From the exact result in the Appendix, we have%
\begin{equation}
R_{\mathrm{c}}=\left\{
\begin{array}{cc}
1/2, & \mu ,\nu >1,\text{or }\mu ,\nu <1 \\
1/2+\eta (\mu ,\nu ), & \text{otherwise}%
\end{array}%
\right.
\end{equation}%
where $\eta (\mu ,\nu )$\ is a nonzero function. On the other hand, it is
presumable that the non-analyticity of $g(\mu ,\nu )$\ at $\mu \nu =1$\ can
result in the non-analyticity of $R_{\mathrm{c}}$. Thirdly, we investigate
the average staggered current, which is defined as%
\begin{equation}
\mathcal{J}=-\frac{2i}{N}\sum\limits_{n=1}^{N/2}(-1)^{n}\langle
\sum\limits_{l=2}^{N}\left( c_{l}^{\dag }c_{l+1}-\mathrm{H.c.}\right)
\rangle _{n}.
\end{equation}%
The feature of non-steady eigen states may also lead to non-analyticity of $%
\mathcal{J}(\mu ,\nu )$ at the non-analytical point of $g(\mu ,\nu )$.
Actually we have%
\begin{equation}
\mathcal{J}=\left\{
\begin{array}{cc}
0, & \mu ,\nu \geqslant 1,\text{or }\mu ,\nu \leqslant 1 \\
\neq 0, & \text{otherwise} \\
4/\pi , & \mu \nu =1%
\end{array}%
\right. .
\end{equation}%
To demonstrate this point, we compute the quantities $\partial
^{n}E_{g}/\partial \theta ^{n}$, $\partial ^{n}R_{\mathrm{c}}/\partial
\theta ^{n}$, and $\partial ^{n}\mathcal{J}/\partial \theta ^{n}$ along the
circle%
\begin{equation}
\nu =1+0.9\cos (\theta ),\mu =1+0.9\sin (\theta )
\end{equation}%
with $n=1,2$. In Fig. \ref{fig5}, we plot these quantities from exact
diagonalization results for finite size system. We find that the density of
ground state energy does not display any critical behaviors as we predicted.
It is different from a conventional QPT. It is understandable since the
gound state does not experience a symmetry breaking as a whole, although a
single-particle eigenstate has time-reversal symmetry breaking. However, the
other two quantities exhibit the characteristics of second-order QPT:
first-order derivatives are non-analytical and second-order derivatives are
divergent.

\section{Conclusion}

\label{Summary}

In summary we have proposed a new type of QPT beyond conventional
symmetry-breaking and topological QPTs. It is based on the concept of
semi-localization state, which is a crossover from extended to localized
state, possessing exponentially decay probability distribution. The peculiar
feature is that the decay length is of the order of the size of the system,
rather than fixed as usual localized state.\ We have shown that such a
semi-localized state can be induced by an asymmetrical dimer in a ring
system. Remarkably, we found that a single dimer can result in a macroscopic
amount of complex energy levels with semi-localized states, which determines
the value of some macroscopic observables, such as the CoM and staggered
current of the many-body ground state. Furthermore, the spectral statistics
is non-analytical as asymmetrical hopping strengths vary, resulting in a
sudden charge of the ground state, i.e., QPT. Another distinguishing feature
of such a QPT is the groundstate energy is analytical at the phase boundary.
The symmetry of the many-body ground state remains unchanged, while
single-particle eigenstate breaks the time-reversal symmetry, resulting the
formation of semi-localized state. It seems that such a quantum phase is
exclusive for non-Hermitian system.

\section*{Acknowledgement}

We acknowledge the support of NSFC (Grants No. 11874225).

\section*{Appendix}

In this Appendix, we present the detailed derivation and analysis for the
Bethe Ansatz solution of the Hamiltonian $H$.

\subsection{Wave function}

We consider the single-particle eigen state%
\begin{equation}
\left\vert \psi _{n}\right\rangle =\frac{1}{\sqrt{\Omega _{n}}}%
\sum_{l=1}^{N}f_{n}^{l}\left\vert l\right\rangle ,
\end{equation}%
following a Bethe Ansatz form%
\begin{equation}
f_{n}^{l}=\left\{
\begin{array}{cc}
1, & l=1 \\
A_{n}e^{ik_{n}l}+B_{n}e^{-ik_{n}l}, & l\in \left[ 2,N-1\right]  \\
f_{n}^{N} & l=N%
\end{array}%
\right. ,  \label{fl}
\end{equation}%
where the normalization factor $\Omega _{n}=\sum_{n=1}^{N}|f_{n}^{l}|^{2}$
is determined by the Dirac inner product $\left\langle \psi _{k}\right.
\left\vert \psi _{k}\right\rangle =1$. The Schrodinger equation%
\begin{equation}
H\left\vert \psi _{n}\right\rangle =\varepsilon _{n}\left\vert \psi
_{n}\right\rangle
\end{equation}%
with eigen energy $\varepsilon _{n}$, can be expressed in a explicit form%
\begin{equation}
f_{n}^{l-1}+f_{n}^{l+1}=\varepsilon _{n}f_{n}^{l},  \label{uniform}
\end{equation}%
within the uniform region and%
\begin{equation}
\left\{
\begin{array}{l}
f_{n}^{3}+f_{n}^{1}=\varepsilon _{n}f_{n}^{2} \\
f_{n}^{2}+\nu f_{n}^{N}=\varepsilon _{n}f_{n}^{1} \\
f_{n}^{N-1}+\mu f_{n}^{1}=\varepsilon _{n}f_{n}^{N} \\
f_{n}^{N-2}+f_{n}^{N}=\varepsilon _{n}f_{n}^{N-1}%
\end{array}%
\right. ,  \label{EquatioinSet}
\end{equation}%
around the asymmetric dimer. Substituting Eq. (\ref{fl}) into Eqs. (\ref%
{uniform}) and (\ref{EquatioinSet}), we have%
\begin{equation}
\varepsilon _{n}=2\cos k_{n},
\end{equation}%
and%
\begin{equation}
\left\{
\begin{array}{c}
A_{n}e^{ik_{n}}+B_{n}e^{-ik_{n}}=f_{n}^{1} \\
A_{n}e^{2ik_{n}}+B_{n}e^{-2ik_{n}}=\varepsilon _{n}f_{n}^{1}-\nu f_{n}^{N}
\\
A_{n}e^{ik_{n}\left( N-1\right) }+B_{n}e^{-ik_{n}\left( N-1\right)
}=\varepsilon _{n}f_{N}-\mu f_{1} \\
A_{k}e^{ik_{n}N}+B_{k}e^{-ik_{n}N}=f_{N}%
\end{array}%
\right. ,  \label{ESi}
\end{equation}%
the solution of Eq. (\ref{ESi}) is%
\begin{equation}
\left\{
\begin{array}{c}
A_{n}=\frac{1-\nu e^{-ik_{n}}f_{n}^{N}}{2i\sin k_{n}}=\frac{\mu
-e^{-ik_{n}}f_{n}^{N}}{2i\sin k_{n}e^{ik_{n}N}} \\
B_{n}=\frac{\nu f_{n}^{N}e^{ik_{n}}-1}{2i\sin k_{n}}=\frac{%
e^{ik_{n}}f_{n}^{N}-\mu }{2i\sin k_{n}e^{-ik_{n}N}}%
\end{array}%
\right. .  \label{ABi}
\end{equation}%
We would like to point out that the argument in the sine function can be
complex number. We note that $A_{n}=(B_{n})^{\ast }$\ if $k_{n}$\ is real,
which indicates the reality of the wave function, $f_{n}^{l}=\left(
f_{n}^{l}\right) ^{\ast }$, obeying the time-reversal symmetry. The
existence of non-trivial solution $\left( A_{n},B_{n}\right) $ requires%
\begin{equation}
\left( \mu +\nu \right) \sin k_{n}=\sin \left[ k_{n}\left( 1+N\right) \right]
+\mu \nu \sin \left[ k_{n}\left( 1-N\right) \right] .  \label{TransEq}
\end{equation}%
And the solution of the wave function\ can be obtained from%
\begin{equation}
\left\{
\begin{array}{c}
A_{n}=\frac{\nu -e^{ik_{n}N}-\nu e^{-2ik_{n}}+\mu \nu e^{ik_{n}\left(
N-2\right) }}{2i\sin k_{n}\left( \nu -e^{ik_{n}N}\right) } \\
B_{n}=\frac{1-\mu \nu }{2i\sin k_{n}\left( \nu e^{-ik_{n}N}-1\right) }%
\end{array}%
\right. ,  \label{ABiii}
\end{equation}%
and%
\begin{equation}
f_{n}^{N}=\frac{e^{-ik_{n}}-\mu e^{ik_{n}\left( N-1\right) }}{\nu
-e^{ik_{n}N}}f_{1}.  \label{f1fN}
\end{equation}%
In the following discussion, we use the normalized wave function%
\begin{equation}
f_{n}^{l}=\sin \left( k_{n}l+\alpha _{n}\right) ,l\in \left[ 1,N\right]
\label{Wavefun}
\end{equation}%
by replacing $f_{n}^{l}$ by $\sqrt{\Omega _{n}}f_{n}^{l}$, where the
coefficients are%
\begin{equation}
\tan \alpha _{n}=\frac{c_{n}}{s_{n}},
\end{equation}%
\begin{eqnarray}
s_{n} &=&\nu ^{2}+1-\nu \cos \left( k_{n}N\right) +\nu \cos \left[
k_{n}\left( N+2\right) \right]   \notag \\
&&-p_{n}\cos k_{n}, \\
c_{n} &=&\nu \sin \left( k_{n}N\right) -\nu \sin \left[ k_{n}\left(
N+2\right) \right] +p_{n}\sin k_{n},  \notag
\end{eqnarray}%
with%
\begin{eqnarray}
p_{n} &=&\cos \left[ k_{n}\left( N+1\right) \right] -\mu \nu \cos \left[
k_{n}\left( N-1\right) \right]   \notag \\
&&+\left( \mu +\nu \right) \cos k_{n}.
\end{eqnarray}%
Similarly, the solution of $H^{\dag }$\ can be obtained as the form%
\begin{equation}
\overline{f_{n}^{l}}(\mu ,\nu )=\left[ f_{n}^{l}(\nu ,\mu )\right] ^{\ast }
\end{equation}%
and obey the biorthonormal relation%
\begin{equation}
\sum_{l}\overline{f_{m}^{l}}f_{n}^{l}=\delta _{mn}
\end{equation}%
if a biorthogonal inner product normalization factor is imposed.

Specifically, in the case with $\mu \nu =1$ and $\mu >1>\nu $ (or $\mu
<1<\nu $),\ Eq. (\ref{TransEq}) reduces to%
\begin{equation}
\sin \left[ k_{n}\left( 1+N\right) \right] +\sin \left[ k_{n}\left(
1-N\right) \right] -\left( \mu +\nu \right) \sin k_{n}=0,
\end{equation}%
and furthermore becomes%
\begin{equation}
\cos \left( Nk_{n}\right) =\frac{\mu +\nu }{2}.
\end{equation}%
Since $(\mu +\nu )/2\geqslant 1$, we have%
\begin{equation}
k_{n}=\frac{2n\pi }{N}+i\phi ,
\end{equation}%
where $\phi =(\ln \nu )/N$. Equation (\ref{ABiii}) gives%
\begin{equation}
\left\{
\begin{array}{c}
A_{k}=e^{-ik_{n}} \\
B_{k}=0%
\end{array}%
\right. ,
\end{equation}%
so we have the wave function,%
\begin{equation}
f_{l}=e^{i\frac{2n\pi }{N}l}e^{-\phi l},\text{ }l\in \left[ 1,N\right] .
\end{equation}%
and the Dirac normalization factor%
\begin{equation}
\Omega _{n}=\sum_{l=1}^{N}\left\vert f_{l}\right\vert ^{2}=\frac{1-e^{-2\phi
N}}{e^{2\phi }-1}.
\end{equation}

\subsection{Spectral statistics and phase diagram}

Now we focus on the solution $k_{n}$\ of the transcend equation in Eq. (\ref%
{TransEq}). Without loss of generality, taking%
\begin{equation}
k_{n}=\frac{2n\pi }{N}+\theta _{n}.
\end{equation}%
For the general case ($n\neq N$\ and $N/2$), Eq. (\ref{TransEq}) becomes%
\begin{equation}
\sin \left( \phi _{n}+\theta _{n}N\right) =\sin \Theta _{n}  \label{theta}
\end{equation}%
where%
\begin{equation}
\tan \phi _{n}=\frac{1+\mu \nu }{1-\mu \nu }\tan \frac{2n\pi }{N}
\label{phi_n}
\end{equation}%
and we define%
\begin{equation}
\Theta _{n}=\arcsin \frac{\left( \mu +\nu \right) \sin \frac{2n\pi }{N}}{%
\sqrt{\left( 1+\mu ^{2}\nu ^{2}\right) -2\mu \nu \cos \frac{4n\pi }{N}}}.
\end{equation}%
We notice that $\phi _{n}$\ is always real. Then the complex $k_{n}$ arises
from the complex $\theta _{n}$, leading to%
\begin{equation}
\left\vert \frac{\left( \mu +\nu \right) \sin \frac{2n\pi }{N}}{\sqrt{\left(
1+\mu ^{2}\nu ^{2}\right) -2\mu \nu \cos \frac{4n\pi }{N}}}\right\vert >1,
\end{equation}%
which is reduced to%
\begin{equation}
\sin ^{2}\frac{2n\pi }{N}>\frac{\left( \mu \nu -1\right) ^{2}}{\left( \mu
-\nu \right) ^{2}}.  \label{Critical n}
\end{equation}%
We find that the most fragile energy level is $n=N/4$,\ so the complex
energy levels start to appear if%
\begin{equation}
\left( 1-\nu \right) \left( 1-\mu \right) <0,
\end{equation}%
which is the appearance condition of the complex levels. The most stable
energy level is $n=1$ or $N-1$, so all the $N-2$ energy levels turn to be
complex at%
\begin{equation}
\sin ^{2}\frac{2\pi }{N}=\frac{\left( \mu \nu -1\right) ^{2}}{\left( \mu
-\nu \right) ^{2}}.
\end{equation}%
We define the proportion of the complex level as the ratio
\begin{equation}
g(\mu ,\nu )=\frac{N_{\mathrm{C}}(\mu ,\nu )}{N}
\end{equation}%
where $N_{\mathrm{C}}$\ is number of the complex levels. In large $N$ limit,
$\sin ^{2}(2\pi /N)\rightarrow 0$, then we have%
\begin{equation}
g(\mu ,\nu )=\left\{
\begin{array}{cc}
0, & \mu ,\nu \geqslant 1,\text{or }\mu ,\nu \leqslant 1 \\
(\pi -2k_{c})/\pi , & \mu >1>\nu ,\text{or }\mu <1<\nu \\
1, & \mu \nu =1%
\end{array}%
\right. ,
\end{equation}%
with $k_{c}=|\arcsin [(1-\mu \nu )/(\nu -\mu )]|$. This expression clearly
shows that $g(\mu ,\nu )$\ is non-analytical at three curves $\mu =1$, $\nu
=1$, and $\mu \nu =1$.

At last, for $n=N$ or $N/2$, Eq. (\ref{TransEq}) is reduced to%
\begin{equation}
\tan \theta _{N}=\frac{\left( 1-\mu \nu \right) \sin \left( \theta
_{N}N\right) }{\mu +\nu -\left( 1+\mu \nu \right) \cos \left( \theta
_{N}N\right) }.
\end{equation}%
$\theta _{N}$\textbf{\ }is either real for some configuration of ($\mu $, $%
\nu $) or complex with Re$\theta _{N}=\pi $ ($0$). For the second case, we
take%
\begin{equation}
\theta _{N}=\pi +i\epsilon \text{ or }i\epsilon ,  \label{thetaN}
\end{equation}%
which corresponds to real energy level but localized state. The decay rate
and energy can be obtained from $\epsilon $, which obeys another transcend
equation%
\begin{equation}
\tan \left( i\phi \right) =\frac{\left( \mu \nu -1\right) \sin \left(
Ni\epsilon \right) }{\left( 1+\mu \nu \right) \cos \left( iN\epsilon \right)
-\left( \mu +\nu \right) }.  \label{TransEqN}
\end{equation}

\subsection{Energy levels}

Next we will show that for a fixed $n$,\ Eq. (\ref{theta}) must have a pair
of solution $\left( \theta _{n},\overline{\theta }_{n}\right) $\ leading to
a pair of $(k_{n}$, $\overline{k}_{n})$. We will discuss it in the following
cases.

(i) Real energy levels. In this case we have%
\begin{equation}
\left\{
\begin{array}{c}
\theta _{n}N=\Theta _{n}-\phi _{n} \\
\overline{\theta }_{n}N=\pi -\Theta _{n}-\phi _{n}%
\end{array}%
\right. ,
\end{equation}%
and accordingly%
\begin{equation}
k_{n}=\frac{2n\pi }{N}+\theta _{n},\overline{k}_{n}=\frac{2n\pi }{N}+%
\overline{\theta }_{n},
\end{equation}%
for $n<N/2$. On the other hand, for the energy level $N-n$, we have%
\begin{equation}
\tan \phi _{N-n}=-\frac{1+\mu \nu }{1-\mu \nu }\tan \frac{2n\pi }{N},
\end{equation}%
which means%
\begin{equation}
\phi _{N-n}=2\pi -\phi _{n},
\end{equation}%
in comparison with Eq. (\ref{phi_n}). Furthermore, from%
\begin{equation}
\sin \left( \phi _{N-n}+\theta _{N-n}N\right) =\sin \Theta _{N-n},
\end{equation}%
we get $\theta _{N-n}$ and $\overline{\theta }_{N-n}$ in the form of%
\begin{equation}
\left\{
\begin{array}{c}
\theta _{N-n}N=\pi -\Theta _{N-n}-\phi _{N-n}=-\overline{\theta }_{n}N \\
\overline{\theta }_{N-n}N=2\pi +\Theta _{N-n}-\phi _{N-n}=-\theta _{n}N%
\end{array}%
\right.
\end{equation}%
and%
\begin{equation}
\left\{
\begin{array}{c}
k_{N-n}=2\pi -\overline{k}_{n} \\
\overline{k}_{N-n}=2\pi -k_{n}%
\end{array}%
\right. .
\end{equation}%
The corresponding energy levels satisfy $\varepsilon _{N-n}=\overline{%
\varepsilon }_{n}=2\cos \overline{k}_{n}$ and $\overline{\varepsilon }%
_{N-n}=\varepsilon _{n}=2\cos k_{n}$. In summary, if $k_{n}$\ ($n<\frac{N}{2}
$) is real and
\begin{equation}
k_{n}=\frac{1}{N}\left[ 2n\pi +\Theta _{n}-\phi _{n}\right] ,
\end{equation}%
with $\varepsilon _{n}$ there must exist another%
\begin{equation}
k_{n^{\prime }}=\frac{1}{N}\left[ (2n^{\prime }+1)\pi -\Theta _{n^{\prime
}}-\phi _{n^{\prime }}\right] ,
\end{equation}%
with $n^{\prime }=N-n$ and $\varepsilon _{n^{\prime }}=\overline{\varepsilon
}_{n}$. We see that $k_{n}$\ is monotonic function except at the point $\mu
=\nu =1$. Thus the real energy levels are non-degeneracy. The corresponding
eigenstate has time reversal symmetry since the wave function is real.

(ii) Complex energy levels. In this case, we have $\sin (\phi _{n}+\theta
_{n}N)>1$. The reality of $\sin (\phi _{n}+\theta _{n}N)$ requires that $%
\theta _{n}$ must be complex since $\phi _{n}$ is real. The two solutions of
Eq. (\ref{theta}) are%
\begin{equation}
\left\{
\begin{array}{c}
\theta _{n}N=\Theta _{n}-\phi _{n} \\
\overline{\theta }_{n}N=(\Theta _{n})^{\ast }-\phi _{n}%
\end{array}%
\right. ,
\end{equation}%
and accordingly%
\begin{equation}
\left\{
\begin{array}{c}
k_{n}=\frac{2n\pi }{N}+\theta _{n} \\
\overline{k}_{n}=\frac{2n\pi }{N}+\overline{\theta }_{n}%
\end{array}%
\right. ,
\end{equation}%
the corresponding energy is $\varepsilon _{n}=2\cos (2n\pi /N+\theta _{n})$
and $\overline{\varepsilon }_{n}=2\cos (2n\pi /N+\overline{\theta }_{n})$.
It indicates that%
\begin{equation}
\overline{\varepsilon }_{n}=\left( \varepsilon _{n}\right) ^{\ast }
\end{equation}%
i.e., the complex energy levels always come in pair. And two energy levels
coalesce when $\theta _{n}=\overline{\theta }_{n}$.\ On the other hand, for
the energy level $N-n$, we have%
\begin{equation}
\phi _{N-n}=2\pi -\phi _{n}
\end{equation}%
which leads to%
\begin{equation}
\left\{
\begin{array}{c}
\theta _{N-n}N=\Theta _{N-n}-\phi _{N-n}=-\theta _{n}N-2\pi  \\
\overline{\theta }_{N-n}N=(\Theta _{N-n})^{\ast }-\phi _{N-n}=-\overline{%
\theta }_{n}N-2\pi
\end{array}%
\right. ,
\end{equation}%
and%
\begin{equation}
\left\{
\begin{array}{c}
k_{N-n}=2\pi -k_{n}-\frac{2\pi }{N} \\
\overline{k}_{N-n}=2\pi -\overline{k}_{n}-\frac{2\pi }{N}%
\end{array}%
\right. .
\end{equation}%
It indicates that the corresponding energy levels obey\textbf{\ }\textrm{Im}$%
k_{N-n}=\mathrm{Im}k_{n}=-\mathrm{Im}\overline{k}_{n}$\textbf{, }$%
\varepsilon _{N-n}\approx \varepsilon _{n}$, $\overline{\varepsilon }%
_{N-n}\approx \overline{\varepsilon }_{n}$\textbf{\ }for large\textbf{\ }$N$
limit\textbf{.}

In summary, if $k_{n}$\ ($n<\frac{N}{2}$) is complex and%
\begin{equation}
k_{n}=\frac{1}{N}\left[ 2n\pi +\Theta _{n}-\phi _{n}\right] ,
\end{equation}%
with $\varepsilon _{n}$ there must exist another%
\begin{equation}
k_{n^{\prime }}=\frac{1}{N}\left[ 2n^{\prime }\pi +(\Theta _{n^{\prime
}})^{\ast }-\phi _{n^{\prime }}\right] ,
\end{equation}%
with $n^{\prime }=N-n$ and $\varepsilon _{n^{\prime }}=\overline{\varepsilon
}_{n}$. We note that the corresponding eigen state breaks time reversal
symmetry since the wave function is complex.

\subsection{Center of mass}

We still estimate the CoM in the following cases.

(i) Real energy levels. In this case, the eigenstate with real $k_{n}$ has
the form%
\begin{equation}
\left\vert \psi _{\mathrm{R}}^{n}\right\rangle =\frac{1}{\sqrt{\Omega _{n}}}%
\sum_{l=1}^{N}\sin \left( k_{n}l+\alpha _{n}\right) \left\vert
l\right\rangle ,
\end{equation}%
where%
\begin{equation*}
\alpha _{n}=\tan \frac{c_{n}}{s_{n}}
\end{equation*}%
and $\Omega _{n}=\sum_{l=1}^{N}\sin ^{2}\left( k_{n}l+\alpha _{n}\right) $
is Dirac normalization factor. Then the CoM of eigenstate $\left\vert \psi _{%
\mathrm{R}}^{n}\right\rangle $\ is%
\begin{eqnarray}
\left\langle r_{c}^{\mathrm{R}}\right\rangle _{n} &=&\frac{1}{N}%
\sum\limits_{l=1}^{N}l\left\langle \psi _{\mathrm{R}}^{n}\right\vert
c_{l}^{\dag }c_{l}\left\vert \psi _{\mathrm{R}}^{n}\right\rangle   \notag \\
&=&\frac{1}{N\Omega _{n}}\sum_{l=1}^{N}l\sin ^{2}\left( lk_{n}+\alpha
_{n}\right) .
\end{eqnarray}%
Taking the approximation $k_{n}\approx 2n\pi /N$, together with the
identities%
\begin{equation}
\left\{
\begin{array}{c}
\sum_{l=1}^{N}l\sin ^{2}\left( lk_{n}+\alpha _{n}\right) \approx \frac{N^{2}%
}{4}-\frac{N\sin \left( k_{n}+2\alpha _{n}\right) }{4\sin k_{n}} \\
\sum_{l=1}^{N}\sin ^{2}\left( lk_{n}+\alpha _{n}\right) \approx \frac{N}{2}-%
\frac{\cos \left( \left( 1+N\right) k_{n}+2\alpha _{n}\right) }{2\sin
k_{n}\csc \left( k_{n}N\right) }%
\end{array}%
\right. ,
\end{equation}%
we have%
\begin{equation}
\left\langle r_{c}^{\mathrm{R}}\right\rangle _{n}\approx \frac{1}{2},
\end{equation}%
which shows that all $\left\vert \psi _{\mathrm{R}}^{n}\right\rangle $ have
the same CoM, locating at the center of the lattice.

(ii) Complex energy levels. In this case, the eigenstates of conjugate pair
are expressed as%
\begin{eqnarray}
\left\vert \psi _{\mathrm{+}}^{n}\right\rangle  &=&\frac{1}{\sqrt{\Omega _{n}%
}}\sum_{l=1}^{N}\sin \left( k_{n}l+\alpha _{n}\right) \left\vert
l\right\rangle , \\
\left\vert \psi _{\mathrm{-}}^{n}\right\rangle  &=&\frac{1}{\sqrt{\Omega _{n}%
}}\sum_{l=1}^{N}\sin \left( k_{n}^{\ast }l+\alpha _{n}^{\ast }\right)
\left\vert l\right\rangle .
\end{eqnarray}%
Similarly, the corresponding CoMs, defined as
\begin{equation}
\left\langle r_{c}^{\mathrm{\pm }}\right\rangle _{n}=\frac{1}{N}%
\sum\limits_{l=1}^{N}l\left\langle \psi _{\mathrm{\pm }}\right\vert
c_{l}^{\dag }c_{l}\left\vert \psi _{\mathrm{\pm }}\right\rangle
\end{equation}%
are identical with each other%
\begin{equation}
\left\langle r_{c}^{\mathrm{+}}\right\rangle _{n}=\left\langle r_{c}^{%
\mathrm{-}}\right\rangle _{n}=\left\langle r_{c}\right\rangle _{n}
\end{equation}%
since $\left\vert \psi _{\mathrm{+}}^{n}\right\rangle \ $and $\left\vert
\psi _{\mathrm{-}}^{n}\right\rangle \ $have\ the same distributions of Dirac
probability. According to Eq. (\ref{Wavefun}), we have%
\begin{eqnarray}
&&\sum_{l=1}^{N}\left\vert \sin \left( k_{n}l+\alpha _{n}\right) \right\vert
^{2}  \notag \\
&\approx &\frac{1}{2}\cosh (k_{\text{\textrm{I}}}+k_{\text{\textrm{I}}%
}N+2\alpha _{\text{\textrm{I}}})\text{\textrm{csch}}(k_{\text{\textrm{I}}%
})\sinh (k_{\text{\textrm{I}}}N)
\end{eqnarray}%
and%
\begin{eqnarray}
&&\sum_{l=1}^{N}l\left\vert \sin \left( k_{n}l+\alpha _{n}\right)
\right\vert ^{2}  \notag \\
&\approx &\frac{1}{8}\text{\textrm{csch}}^{2}k_{n}^{\text{\textrm{I}}}(\cosh
2\alpha _{n}^{\text{\textrm{I}}}-\left( 1+N\right) \cosh [2(k_{n}^{\text{%
\textrm{I}}}N+\alpha _{n}^{\text{\textrm{I}}})]  \notag \\
&&+N\cosh [2(k_{n}^{\text{\textrm{I}}}+k_{n}^{\text{\textrm{I}}}N+\alpha
_{n}^{\text{\textrm{I}}})]),
\end{eqnarray}%
where%
\begin{eqnarray}
k_{n}^{\text{\textrm{R}}} &=&\text{\textrm{Re}}k_{n}\text{, }k_{n}^{\text{%
\textrm{I}}}=\text{\textrm{Im}}k_{n}\text{,} \\
\alpha _{n}^{\text{\textrm{R}}} &=&\text{\textrm{Re}}\alpha _{n}\text{, }%
\alpha _{n}^{\text{\textrm{I}}}=\text{\textrm{Im}}\alpha _{n}\text{.}  \notag
\end{eqnarray}%
Finally we get%
\begin{widetext}
\begin{equation}
\left\langle r_{c}\right\rangle _{n}\approx \frac{N^{-1}\cosh \left( 2\alpha
_{n}^{\text{\textrm{I}}}\right) -\cosh [2(k_{n}^{\text{\textrm{I}}}N+\alpha
_{n}^{\text{\textrm{I}}})]+\cosh [2(k_{n}^{\text{\textrm{I}}}+k_{n}^{\text{%
\textrm{I}}}N+\alpha _{n}^{\text{\textrm{I}}})]}{4\cosh (k_{n}^{\text{%
\textrm{I}}}+k_{n}^{\text{\textrm{I}}}N+2\alpha _{n}^{\text{\textrm{I}}%
})\sinh (k_{n}^{\text{\textrm{I}}}N)\sinh (k_{n}^{\text{\textrm{I}}})}
\end{equation}%
\end{widetext}which indicates that the CoM of complex level has distribution
from $0$\ to $1$.

For the special case with $\mu \nu =1$, and $\mu >1>\nu $ $($or $\mu <1<\nu
) $, it readily to obtain%
\begin{equation}
\left\langle r_{c}\right\rangle _{n}\approx \frac{1}{(\sqrt[N]{\nu ^{2}}-1)N}%
+\frac{1}{1-\nu ^{2}}
\end{equation}%
in large $N$ limit, which is independent of $n$.

\subsection{Current}

We now turn to the current of eigenstate, which is defined as%
\begin{eqnarray}
J_{l}^{n} &=&-i\left\langle \left( c_{l}^{\dagger }c_{l+1}-\mathrm{H.c.}%
\right) \right\rangle _{n}  \label{Jnl} \\
&=&-i(\left( f_{n}^{l}\right) ^{\ast }f_{n}^{l+1}-\left( f_{n}^{l+1}\right)
^{\ast }f_{n}^{l}).  \notag
\end{eqnarray}%
According to Eq. (\ref{Wavefun}), for the eigenstates with real $k_{n}$, we
always have%
\begin{equation}
J_{l}^{n}=0.
\end{equation}%
In contrast, for the eigenstates with complex $k_{n}$, we have%
\begin{eqnarray}
J_{l}^{n} &=&-i(\sin \left( k_{n}^{\ast }l+\alpha _{n}^{\ast }\right) \sin
\left( k_{n}l+\alpha _{n}+k_{n}\right)  \\
&&-\sin \left( k_{n}^{\ast }l+k_{n}^{\ast }+\alpha _{n}^{\ast }\right) \sin
\left( k_{n}l+\alpha _{n}\right) ).  \notag
\end{eqnarray}%
Taking a trigonometric transformation and an approximation $\sinh (k_{n}^{%
\text{\textrm{I}}})\approx 0$, one can obtain%
\begin{equation}
J_{l}^{n}\approx -\sin (k_{n}^{\text{\textrm{R}}})\sinh (2k_{n}^{\text{%
\textrm{I}}}l+2\alpha _{n}^{\text{\textrm{I}}}+k_{n}^{\text{\textrm{I}}}).
\notag
\end{equation}%
We see that the current with $k_{n}^{\ast }$ is $\left( J_{l}^{n}\right)
^{\ast }=-J_{l}^{n}$, i.e., the sum current of a conjugate pair always
vanishes. We introduce the concept of the average staggered current,%
\begin{equation}
\mathcal{J}=-\frac{2i}{N}\sum\limits_{n=1}^{N/2}(-1)^{n}\langle
\sum\limits_{l=2}^{N}\left( c_{l}^{\dag }c_{l+1}-\mathrm{H.c.}\right)
\rangle _{n},
\end{equation}%
which is nonzero for the band containing complex levels. A direct derivation
yields%
\begin{eqnarray}
&&\langle \sum\limits_{l=2}^{N}\left( c_{l}^{\dag }c_{l+1}-\mathrm{H.c.}%
\right) \rangle _{n} \\
&=&-i\sin (k_{n}^{\text{\textrm{R}}})\text{\textrm{csch}}(k_{n}^{\text{%
\textrm{I}}})\sinh (k_{n}^{\text{\textrm{I}}}N)\sinh (2k_{n}^{\text{\textrm{I%
}}}+2\alpha _{n}^{\text{\textrm{I}}}+Nk_{n}^{\text{\textrm{I}}}).  \notag
\end{eqnarray}%
for large $N$. The average staggered current has the from%
\begin{eqnarray}
\mathcal{J} &=&-\frac{2}{N}\sum\limits_{n=1}^{N/2}(-1)^{n}\sin (k_{n}^{\text{%
\textrm{R}}})\text{\textrm{csch}}(k_{n}^{\text{\textrm{I}}})  \notag \\
&&\times \sinh (k_{n}^{\text{\textrm{I}}}N)\sinh (2k_{n}^{\text{\textrm{I}}%
}+2\alpha _{n}^{\text{\textrm{I}}}+Nk_{n}^{\text{\textrm{I}}}).
\end{eqnarray}%
For the special case with $\mu \nu =1$, in large $N$\ limit, it readily to
obtain%
\begin{equation}
\mathcal{J}\approx \frac{4}{\pi }.
\end{equation}%
In summary, we have%
\begin{equation}
\mathcal{J}=\left\{
\begin{array}{cc}
0, & \mu ,\nu \geqslant 1,\text{or }\mu ,\nu \leqslant 1 \\
\neq 0, & \text{otherwise} \\
4/\pi , & \mu \nu =1%
\end{array}%
\right. ,
\end{equation}%
which has the implication that $\mathcal{J}$\ can characterize the phase
transitions.

\end{document}